\documentclass[journal]{IEEEtranTIE}
\usepackage{graphicx}
\usepackage{cite}
\usepackage{picinpar}
\usepackage{amsmath}
\usepackage{url}
\usepackage{flushend}
\usepackage[latin1]{inputenc}
\usepackage{colortbl}
\usepackage{soul}
\usepackage{multirow}
\usepackage{pifont}
\usepackage{color}
\usepackage{alltt}
\usepackage[hidelinks]{hyperref}
\usepackage{enumerate}
\usepackage{siunitx}
\usepackage{breakurl}
\usepackage{epstopdf}
\usepackage{pbox}

\usepackage{cite}

\usepackage{graphicx}

\usepackage{amsmath}
\usepackage{amsfonts} 

\usepackage[tight,footnotesize]{subfigure}
\usepackage{gensymb}

\usepackage{glossaries}

\begin{document}
\title{Designing a Pseudo-Random Bit Generator with a Novel 5D-Hyperchaotic System}

\author{
	\vskip 1em
	
	Ngoc T. Nguyen,
	Toan Q. Bui, 
	Ghyslain Gagnon, \emph{Member, IEEE}, 
	Pascal Giard, \emph{Senior Member, IEEE}, 
	\\ and Georges Kaddoum, \emph{Member, IEEE}, 

	\thanks{
	
	The authors are with the LaCIME, Department of Electrical Engineering, \'Ecole de technologie sup\'erieure, Montreal,
	QC, Canada, (e-mails: ngoc.nguyen-thi-thu.1@ens.etsmtl.ca, quang-the-toan.bui.1@ens.etsmtl.ca, ghyslain.gagnon@etsmtl.ca, pascal.giard@etsmtl.ca, georges.kaddoum@etsmtl.ca).
	
	}
}

\newacronym{fpga}{FPGA}{field-programmable gate array}
\newacronym{ff}{FF}{flip-flop}
\newacronym{lfsr}{LFSR}{linear-feedback shift register}
\newacronym{lut}{LUT}{look-up table}
\newacronym{prng}{PRNG}{pseudo-random number generator}
\newacronym{ffrk}{FFRK}{fourth-folding Runge-Kutta}

\maketitle
	
\begin{abstract}
Dynamic and non-linear systems are emerging as potential candidates for random bit generation. In this context, chaotic systems, which are both dynamic and stochastic, are particularly suitable. This paper introduces a new continuous chaotic system along with its corresponding implementation, which targets \glspl{fpga}. This chaotic system has five dimensions, which exhibit complex chaotic dynamics, thus enabling the utilization of chaotic signals in cryptography. A mathematical analysis is presented to demonstrate the dynamic characteristics of the proposed hyperchaotic system. A novel digital implementation of the proposed system is presented. Moreover, a data scrambling circuit is implemented to eliminate the bias effect and increase the randomness of the bitstream generated from the chaotic signals. We show that the proposed random bit generator has high randomness. The generated bits successfully pass well-known statistical randomness test-suites, i.e., NIST SP800-22, Diehard and TestU01. The ready-to-use random bit generator is deployed on a Xilinx Zynq-7000 SoC ZC702 Evaluation Kit. Experimental results show that the proposed random bit generator can achieve a maximum throughput of 6.78\,Gbps, which is over 3.6 times greater than state-of-the-art designs, while requiring under 4\% of the resources available on the targeted \gls{fpga}.
\end{abstract}

\glsresetall


{}

\definecolor{limegreen}{rgb}{0.2, 0.8, 0.2}
\definecolor{forestgreen}{rgb}{0.13, 0.55, 0.13}
\definecolor{greenhtml}{rgb}{0.0, 0.5, 0.0}

\definecolor{revcolor}{rgb}{1, 0., 0.} 

\section{Introduction}

\IEEEPARstart{R}{andom} number generators are critical components that are responsible for generating public keys, private keys, and other kinds of random numbers that are utilized in cryptographic applications and security \cite{li_novel_2019, liu_cryptanalysis_2020,moysis_chaos_2020}.  
Differential chaotic systems, which present high dynamic characteristics and multi-dimensional signals, are superior in terms of generating random bits due to their ability to achieve a high level of randomness \cite{NN-Access}. Therefore, in this work, we focus on the design of such systems, targeted at \glspl{fpga}. 

A hyperchaotic system exhibits rich dynamics since the system states it hosts are expanded exponentially in several directions simultaneously. This property makes the hyperchaotic system an interesting candidate for the generation of random keys used in miscellaneous applications in engineering, such as secure communications, cryptosystems, and encryptions \cite{liu_hyperchaotic_2016, irfan_pseudorandom_2020}. Therefore, in the present work, we develop a 5D hyperchaotic system with three positive Lyapunov exponents to provide better dynamic characteristics than the state of the art. Moreover, high-dimensional chaotic systems provide multiple outputs which improve the throughput of the overall random bit generator. 

The implementation methodology adopted has a great impact on the digitalization of differential chaotic systems. The research in \cite{zidan_effect_2011} presented an implementation of numerical techniques, including the Euler, mid-point, and fourth-order Runge-Kutta (RK) methods on \glspl{fpga}. Among these, the Euler method has the shortest data path, but the least accuracy. The fourth-order RK method shows the highest accuracy, but also has the longest time frame. 
In state-of-the-art RK algorithm implementations, mapping functions are used four times and implemented separately \cite{akgul_chaos-based_2016,koyuncu_design_2017}. We propose the \gls{ffrk} method to allow reusing the mapping functions (the latter are implemented only once and reused four times), and as a result, the device resources required to implement the mapping functions are reduced by 75\%. The iterations are controlled by adding MUXs and control signals, which use insignificant amounts of device resources. Moreover, taking advantage of the multiple dimensional signals in hyperchaotic systems, we implement an effective post-processing, in which five chaotic outputs are used to generate ready-to-use random bitstreams. 
In summary, the contributions of this work include i) a new 5D hyperchaotic system to generate stochastic signals, ii) a novel implementation of the fourth-order RK algorithm, and iii) a simple and effective data post-processing circuit. 

The remainder of this paper is organized as follows. Related works are reviewed in Section~\ref{sec: related}. The mathematical model of the hyperchaotic system is clarified in Section~\ref{sec:5d-syst-model}. The implementation of the proposed hyperchaotic system in a \gls{fpga} hardware device using the new \gls{ffrk} method is detailed in Section~\ref{sec:prng-scheme}. The experimental resuts are presented and discussed in Section~\ref{sec:exp}. Then, the randomness evaluation of the random bit generator is presented in Section~\ref{sec:perf-eval}. Finally, Section~\ref{sec:conclusion} summarizes the work in this paper.

\section{Related Works} \label{sec: related}
In this section, we review the differential chaotic systems and their implementations in the digital world to emphasize our research contribution. There are several commonly used differential chaotic systems in secure communications, including the Lorenz system, Chua's system, Liu's system and Lu's system \cite{ zhang_system_2017, tolba_fpga_2017}. The Lorenz system has been used in different applications in many research works \cite{zhang_system_2017}. The research in \cite{zhang_system_2017} presents the implementation of a Lorenz system in \gls{fpga} hardware devices and co-simulation with Matlab. Liu's system is implemented in \cite{tolba_fpga_2017} using the Grunward-Letniknov algorithm. However, the above chaotic systems utilize multiple multiplexers, which are resource-demanding in hardware implementations. Moreover, the authors in \cite{koyuncu_analog_2013} provide both analog and digital implementations of the Bruke-Shaw chaotic system in a Virtex-6 \gls{fpga} chip, which consumes a lot of device resources. The research in \cite{tlelo-cuautle_fpga_2015} presents an \gls{fpga} realization of Chua's system with multiple scrolls by using different saturated functions. The work in \cite{akgul_chaos-based_2016} implements a 3D chaotic system with no equilibrium points. A high-speed \gls{fpga} implementation of a 3D continuous chaotic system, which achieves a maximum operating frequency of 293\,MHz, is presented in \cite{koyuncu_design_2017}. 

The authors in \cite{avaroglu_hybrid_2015} present a new hybrid PRNG using Sprott 94 G chaotic system which improves security of the advanced encryption standard (AES) in cryptographic systems.
Although instructive, these chaotic systems are 3-dimensional systems which has only one positive Lyapunov exponent. This means that thay have limited dynamic characteristics as the signal expands exponentially in only one direction. Speed optimization is applied to a 3D chaotic system and then expanded to a 4D chaotic system in \cite{bonny_hardware_2019}, achieving a maximum throughput of 1882\,Mbps for the random bit generator. 
	
Nowadays, differential  chaotic  systems  achieve  higher  complexity  by conveying integer-order systems into the fractional-order domain \cite{ akgul_design_2019}.  However,  fractional-order  systems  are  complicated to implement in hardware design due to their memory dependency. The hardware implementation of fractional-order differentiators and integrators requires careful consideration \cite{tolba_fpga_2017}. 

Although chaotic systems are unpredictable, and have random-like state trajectories, they can be studied and recovered by using computational tools. A cyber-attack has a high possibility of success if the target system uses a well-known and self-excited oscillator for its random bit generator. After the transient process, a trajectory, starting from the point of an unstable manifold in a small neighborhood of unstable equilibrium, can be revealed. Therefore, the system's parameters can be computed to recover the target system. From this point of perspective, the development of modern computers enables the numerical simulation of complex nonlinear dynamical systems, and therefore the structure of their trajectories can be deduced. However, this approach shows very limited success when it comes to hidden attractor chaotic systems \cite{barati_simple_2016, fozin_fonzin_coexisting_2018}. The hyperchaotic systems with hidden attractors in \cite{bao_initial_2018, cavusoglu_new_2019} were proposed to overcome such attacks. As compared to previous systems, a mathematical analysis of our hyperchaotic system with hidden attractors shows rich dynamic characteristics that are suitable for use in security and cryptographic applications.
\section{5D Differential Chaotic System Model}\label{sec:5d-syst-model}
The following section presents the proposed hidden attractor hyperchaotic system, which is expressed by five differential equations as given in (1). The theoretical analysis is divided into three parts. The first part presents the proposed chaotic system and a theoretical study of chaotic characteristics, while the second part addresses the stability of the equilibrium. The simulation of chaotic transitions is investigated in the third part. 
\subsection{Mathematical Analysis}
The proposed 5D dynamic system, which is developed from the 4D hyperjerk chaotic system presented in \cite{pham_chaotic_2016}, is expressed as $S'=F(S)$, $S=(x, y, z, u, v)\in\mathbb{R}^5$, with
\begin{equation}
F(S)=
\begin{cases}
x'=y,  \\
y'=z, \\
z'=u, \\
u'=-z-0.5\times u+(x-1)\times y \\
v'=-u-0.5\times v+(x-1)\times z.\
\end{cases}
\label{eq: system}
\end{equation}
\indent The ordinary differential equations (ODEs) are solved and simulated in MATLAB, based on the fourth-order RK integration algorithm with a step size of $10^{-2}$. Here, a small step size is chosen to provide a higher resolution and better accuracy.  
The equilibrium points of the proposed system are located on the line $E(x,0,0,0,0)$. 
Therefore, the proposed system is a dynamical system with hidden attractors. According to \cite{cavusoglu_new_2019}, it is difficult to expose a chaotic attractor, and then reveal the chaotic system architecture by choosing an arbitrary initial condition. In other words, the chaotic attractors are invisible to attackers and their basin localization does not dissolve the chaotic oscillator. In what follows, we present a mathematical analysis of the novel chaotic system in terms of Lyapunov exponents and bifurcation diagrams. 
\begin{enumerate}[]
	\item Lyapunov exponents: The Lyapunov exponents of the system are defined as:  $L_1=\lim_{t\rightarrow \infty}\frac{1}{t} \log\frac{\|\partial x(t)\|}{\|\partial x(0)\|}$, $L_2=\lim_{t\rightarrow \infty}\frac{1}{t} \log\frac{\|\partial y(t)\|}{\|\partial y(0)\|}$, $L_3=\lim_{t\rightarrow \infty}\frac{1}{t} \log\frac{\|\partial z(t)\|}{\|\partial z(0)\|}$, 
	$L_4=\lim_{t\rightarrow \infty}\frac{1}{t} \log\frac{\|\partial u(t)\|}{\|\partial u(0)\|}$,
	and 
	$L_5=\lim_{t\rightarrow \infty}\frac{1}{t} \log\frac{\|\partial v(t)\|}{\|\partial v(0)\|}$
	The Lyapunov exponents of the novel 5D chaotic system are: $L_1=0.093790,L_2=0.001101, L_3=0.000107, L_4=-0.500100$, and $L_5=-0.594898$. The initial conditions of the system are chosen as $x_0=0.0002,y_0=0.0005,z_0=0.00005,v=0.001, \text{and } v_0=0$.
	The divergence of (\ref{eq: system}) is evaluated based on the following conditions: 
	\begin{equation}
	\begin{split}
	\frac{\partial \dot{x}}{\partial x}+\frac{\partial \dot{y}}{\partial y}+
	\frac{\partial \dot{z}}{\partial z}+
	\frac{\partial \dot{u}}{\partial u}+\frac{\partial \dot{v}}{\partial v}=-1 < 0,\\
	L=L_1+L_2+L_3+L_4+L_5 < 0.
	\end{split}
	\end{equation}
	\item Lyapunov dimension: The Lyapunov dimension ($D_L$), which is closely related to the correlation dimension, is commonly used to evaluate the chaotic complexity \cite{chlouverakis_comparison_2005}. A higher Lyapunov dimension value is associated with a chaotic system with a higher level of complexity.  
	\begin{equation}
	D_{L}=j+\frac{1}{|L_{j+1}|}\sum_{i=1}^j L_i,
	\end{equation}
	where $j$ is the largest index of the positive Lyapunov exponent. In the proposed system, $j=3$, and the Lyapunov dimension is therefore: 
	\begin{equation}
	D_{L}=3+\frac{L_1+L_2+L_3}{|L_4|}=3.1899.
	\end{equation}
\end{enumerate}

\begin{figure}
	\includegraphics[width=3.5in]{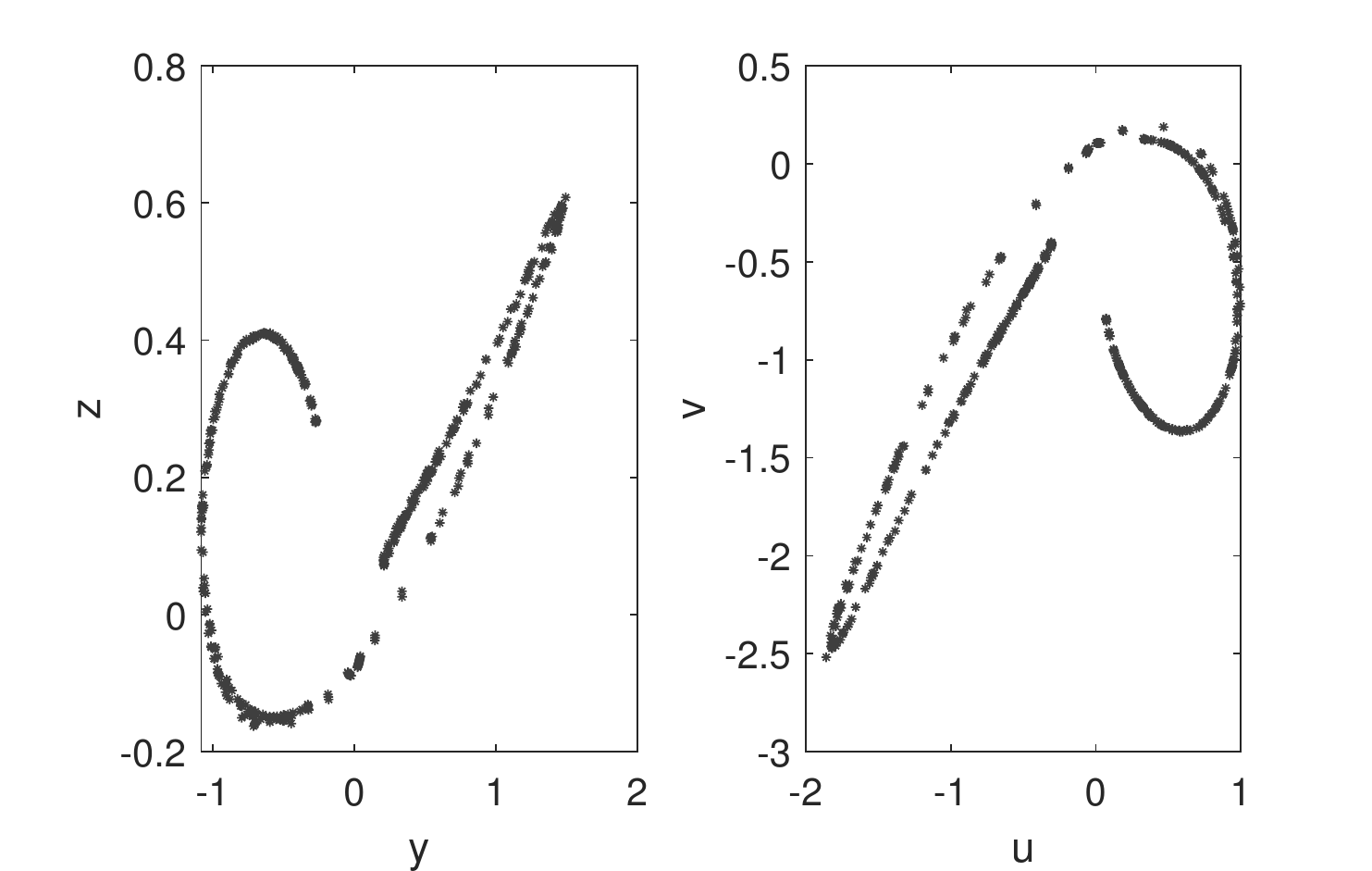}
	\caption{Poincaré section of the phase-space in the plain ($x=1$).}
	\label{fig:Poicare}
\end{figure}

Fig. \ref{fig:Poicare} shows the Poincaré section of the phase-space in the plane $x=1$. Distinct set of points in the Poincaré section indicates the chaotic region of dynamic system.
\subsection{Stability of Equilibria}
The stability evaluation of the equilibria is an important step to find the chaotic region of a dynamic system. To evaluate the stability of equilibria, the Jacobian matrix at the equilibria $E=[c,0,0,0]$ of the proposed hyperchaotic system is calculated. The eigenvalues of the Jacobian matrix satisfy the condition $J_E-\lambda I=0$. 
\begin{equation}
J_E=
\begin{bmatrix}
0	&	1	&	0	&	0	&	0	\\
0	&	0	&	1	&	0	&	0	\\
0	&	0	&	0	&	1	&	0 	\\
0	&	(c-1)	&	-1	&	-0.5 &  0	\\
0	& 	0	&	(c-1)&	-1	&	-0.5	\\
\end{bmatrix}.
\end{equation}
\indent Assuming a small perturbation from the fixed points $x(t)=x(0)+\Delta x$. If $\Delta x \approx e^\lambda t$, the characteristic polynomial equation is derived as:  
\begin{equation}
\lambda(\lambda+0.5)(\lambda^2+1.5\lambda+(c-0.5))=0.
\label{eq:lambda}
\end{equation}
\indent This equation indicates that the Jacobian matrix at equilibrium points has one zero value and three non-zero values. According to the Routh-Hurwitz criterion, apart from the zero eigenvalues, the real parts of the roots of (\ref{eq:lambda}) are negative if and only if $(c-0.5)>0$. To make the equilibrium set $E$ unstable, thereby enabling the possibility of chaos occurrence, the initial condition $x(0)=c$ must satisfy $c<0.5$. In other words, depending on the initial value, the proposed system has stable or unstable saddle-focus points. Thus, the dynamical behaviors of the line equilibrium chaotic system are heavily dependent on the initial state of the variable $x(t)$.
\subsection{Transitions to Chaotic Region}
\begin{figure}
	\centering
	\includegraphics[width=3.5in]{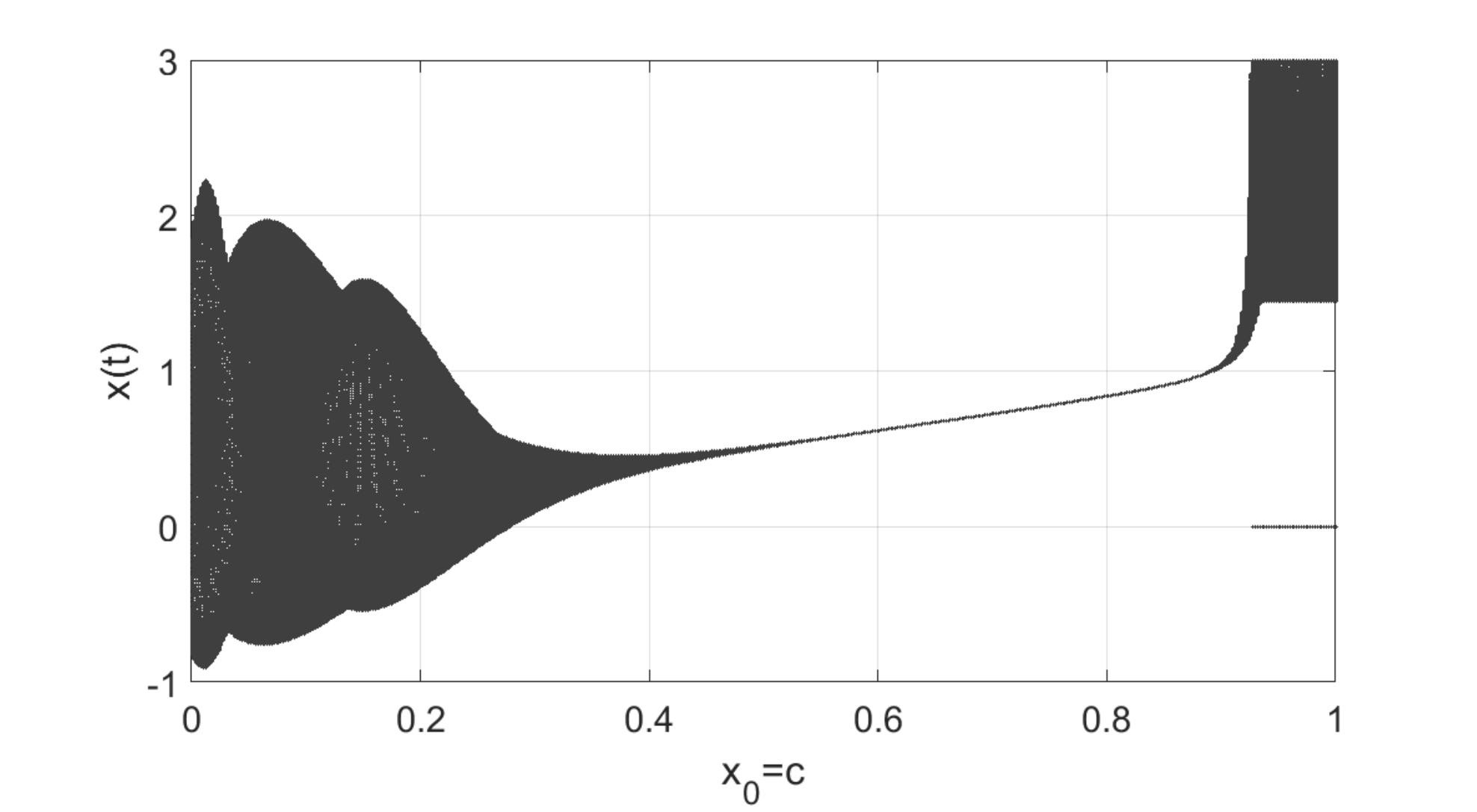}
	\caption{\centering Bifurcation diagram according the initial condition.}
	\label{fig:initialcond}
\end{figure}

\begin{figure}
	\centering
	\includegraphics[width=\linewidth]{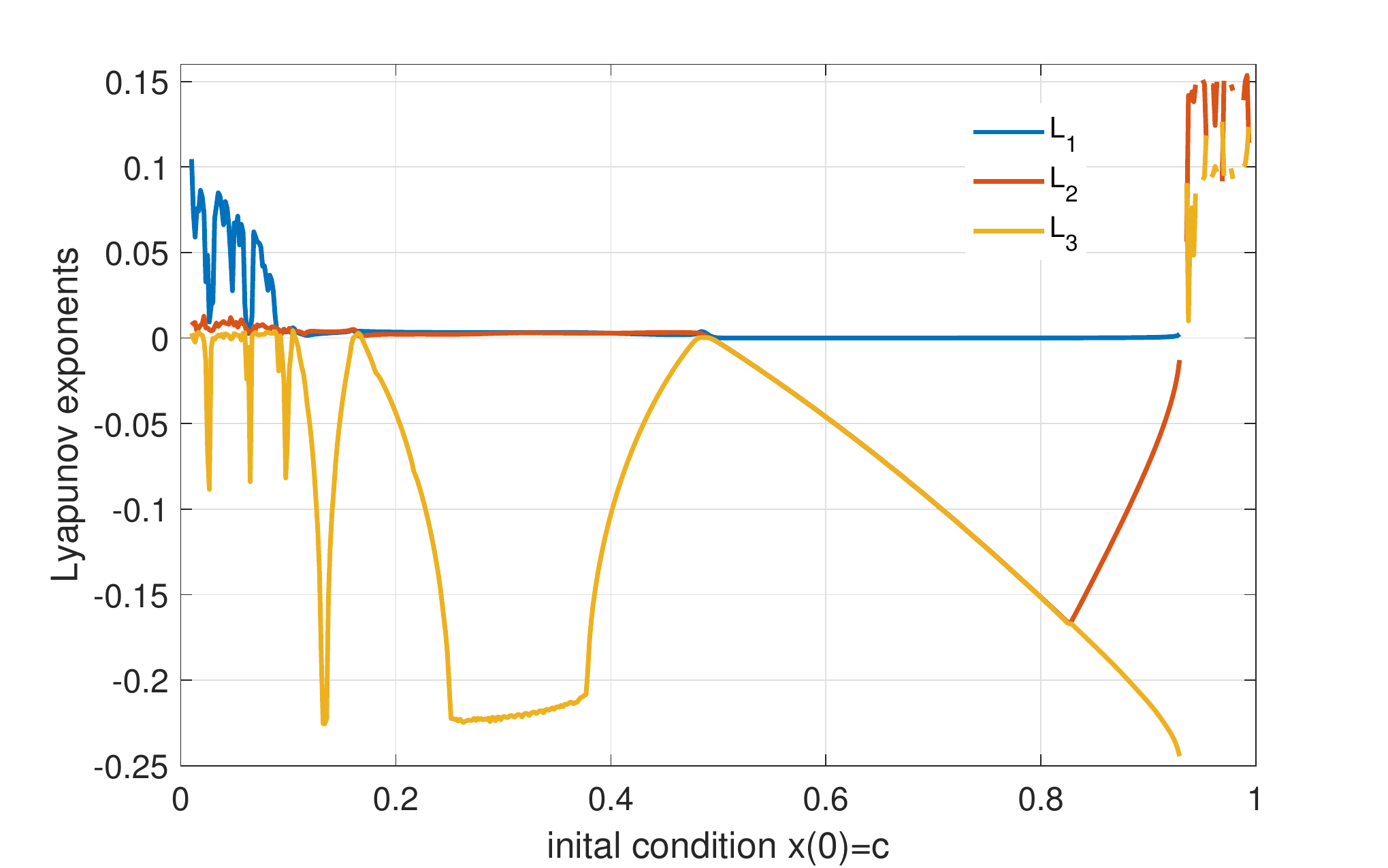}
	\caption{\centering Lyapunov spectrum according to initial condition x(0)=c.}
	\label{fig:LyaSpec}
\end{figure}
\begin{figure*}[!t]
	\centering
	\includegraphics[width=\linewidth]{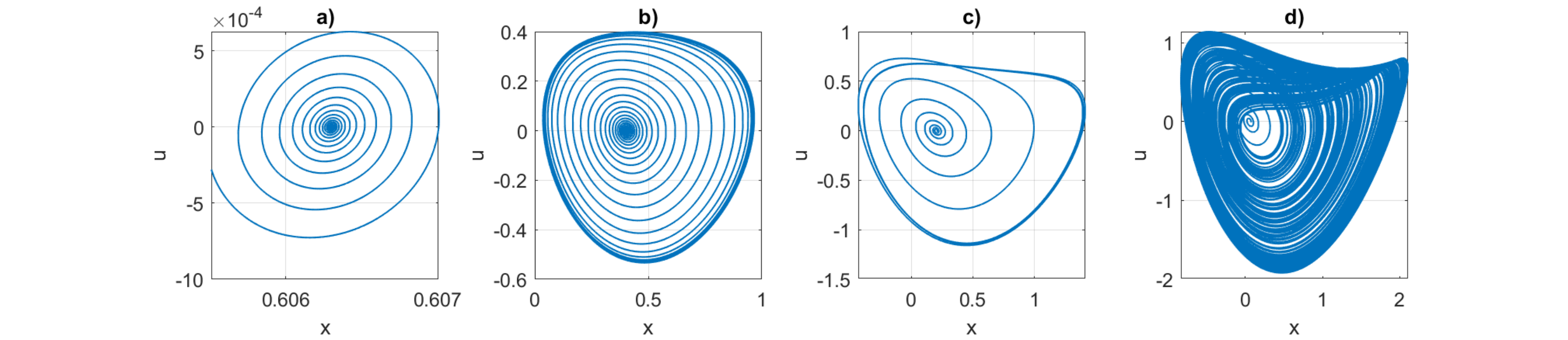}
	\caption{\centering Transition to chaos regions depending to initial condition $x(0)=c$\,: a) stable region for $c=0.6$, b) limited dynamic region for $c=0.4$, c) chaotic region with limited periods for $c=0.2$, and d) rich dynamic and chaotic region for $c=0.05$.}
	\label{fig:ChaosTransition}
\end{figure*}
The bifurcation diagram and the Lyapunov spectrum of the state variable $x(t)$ of the proposed system are shown in Fig.\,\ref{fig:initialcond} and Fig.\,\ref{fig:LyaSpec}, respectively. Both figures illustrate that the value of the initial condition $c$ has an impact on the characteristics of the system. Moreover, a stable region is observed for $0.5<c<0.92$ in Fig.\,\ref{fig:initialcond}, where the data space converges. This corresponds to the negative Lyapunov exponents in Fig.\,\ref{fig:LyaSpec}. A dynamic and unbounded region is observed for $c>0.92$, a region that corresponds to positive high-value Lyapunov exponents in Fig.\,\ref{fig:LyaSpec}. Finally, in the $0<c<0.5$ range, the system is dynamic and bounded in a limited data space, as illustrated in the bifurcation diagram. This corresponds to the presence of positive Lyapunov exponents $L_1$ and/or $L_2$, as shown by the Lyapunov spectrum. However, the system only exhibits rich and chaotic dynamics for $c<0.05$, where a high positive value of $L_1$ is presented, as shown in Fig.\,\ref{fig:LyaSpec}. The gradual transition from periodic orbits to chaotic regions based on various values of the initial condition $c$ is observed in Fig. \,\ref{fig:ChaosTransition}. The chaotic region with rich dynamics illustrated under Fig. \,\ref{fig:ChaosTransition}-d) is obtained with $c=0.05$ which is compatible to the stability analysis of equilibria in the previous section.  
\begin{figure*}[!t]
	\centering
	\includegraphics[width=0.7\linewidth]{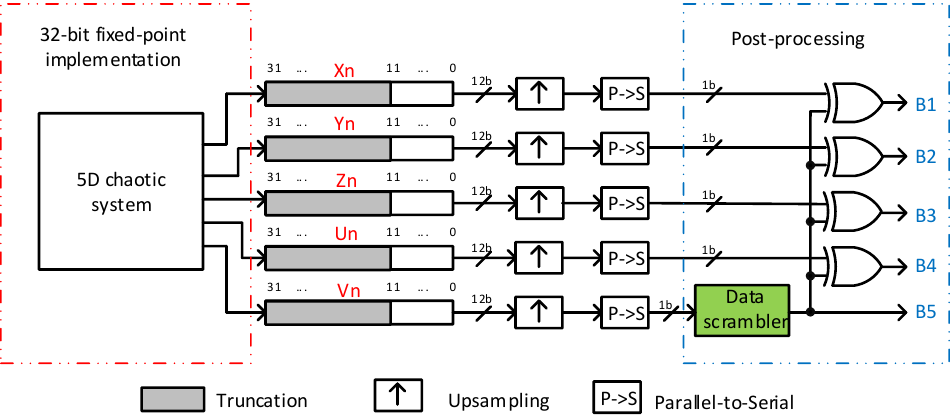}
	\caption{\centering The proposed pseudo-random number generator scheme.}
	\label{fig:PRNGScheme}
\end{figure*}

\section{Pseudo-Random Number Generator Scheme}\label{sec:prng-scheme}
In this section, we present a pseudo-random number generator scheme based on the chaotic output signals. We use fixed-point 32-bits to represent data, with 1-bit for the sign and 27 bits for the fractions. Fig.\,\ref{fig:PRNGScheme} shows the scheme of the proposed random bit generator which includes the implementation of the 5D chaotic system block using \gls{ffrk} with 32-bit fixed-point data. The truncation block that follows keeps the 12 least-significant bits of the stream. Then, the data is up-sampled and converted from parallel to serial before entering the post-processing block. In the post-processing block, a data scrambler is proposed using a simple LFSR of one of the chaotic outputs, while the random bits are obtained by XORing the output of the data scrambler with the other chaotic signals. 

Due to the long data path of the RK algorithm and low self-oscillator frequency of the chaotic system, signal truncation is applied in order to have better randomness qualification in the following block. Therefore, 32-bit chaotic outputs are truncated, where only 12 least significant bits (LSB) are used in the following post-processing. 12-LSBs are up-sampled and serialized before the post-processing step. 
In what follows, we present a novel \gls{fpga}-based implementation of the fourth-order RK algorithm in a new manner to solve the differential chaotic equations. Moreover, we present the post-processing process with a data scrambler to increase the randomness of the generated bitstreams and eliminate bias effects.
\begin{figure*}
	\centering
	\includegraphics[width=0.8\linewidth]{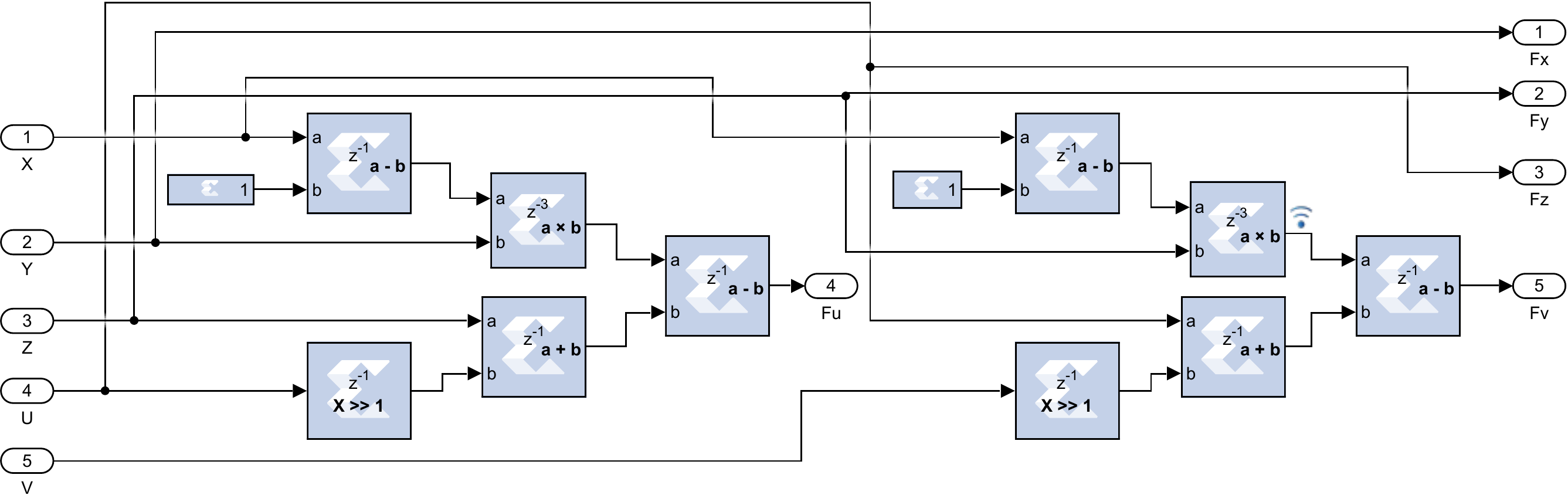}
	\caption{\centering Hyperchaos function $F(S_n)$ implementation in Xilinx System Generator.}
	\label{fig:HyperJerkFunc}
\end{figure*}
\begin{figure*}[!t]\centering
	\includegraphics[width=0.7\linewidth]{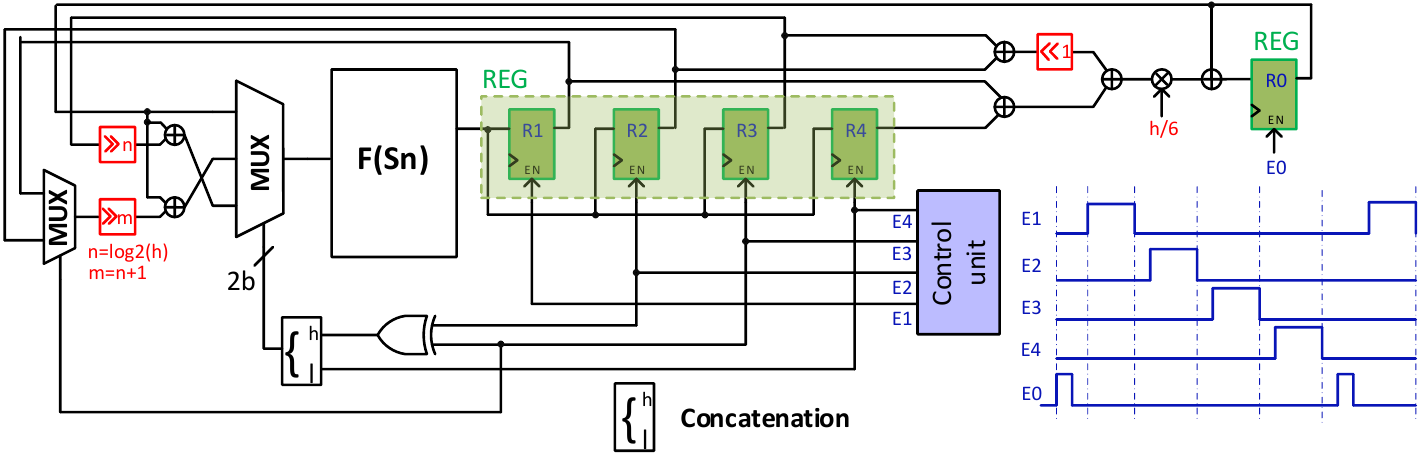}
	\caption{\centering Block diagram of the proposed \glsfirst{ffrk} algorithm. Clock signals are omitted for clarity.}
	\label{fig:FFRK}
\end{figure*}

\begin{figure*}
	\centering
	\includegraphics[width=\linewidth]{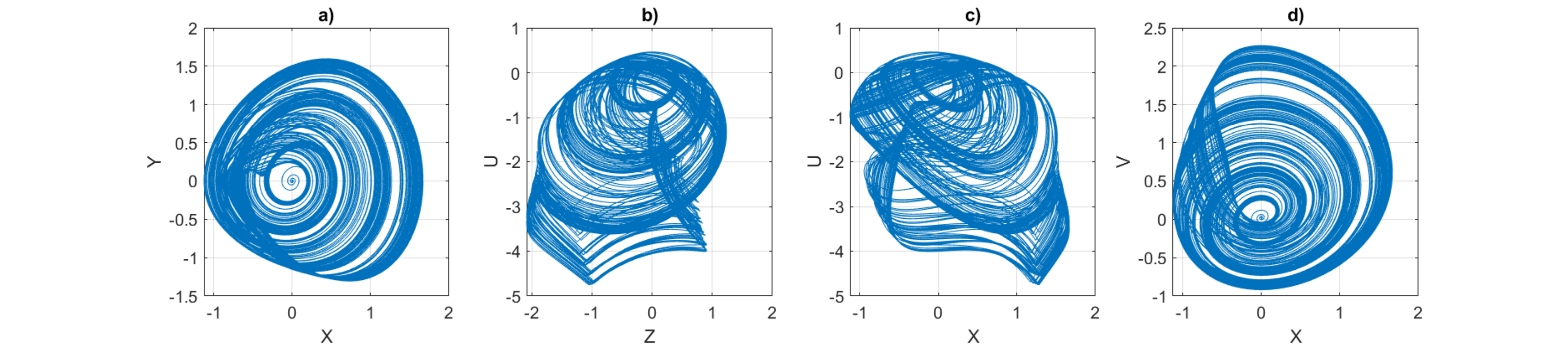}
	\caption{Trajectories of the chaotic outputs.}
	\label{fig:Trajectory}
\end{figure*}
\subsection{Implementation of the 5D Hyperchaotic System}
To solve differential equations, there are three well-known numerical methods, with different complexity and accuracy. These are, the Euler method, the Midpoint method, and the RK method. Compared to the other two candidates, the RK method has the longest calculation path and requires larger areas \cite{zidan_effect_2011}. However, it produces the most accurate results. The proposed \gls{ffrk} implementation optimizes the resources of the 5D hyperchaotic system. Compared to previous RK implementations, the proposed \gls{ffrk} provides the most accurate solution for differential equations and consumes fewer device resources. 
The system state at the $n$-th time step is denoted by the vector $S_n=[x_n,y_n,z_n,u_n,v_n]$. The proposed hyperchaotic system discrete-time implementation is calculated as follows: 
\begin{equation}
F(S_n)=
\begin{cases}
F_x=y_n \\
F_y=z_n \\
F_z=u_n \\
F_u=-z_n-0.5\times u_n+(x_n-1)\times y_n \\
F_v=-u_n-0.5\times v_n+(x_n-1)\times z_n.
\end{cases} \\
\label{eq:hyperchaosfunc}
\end{equation}
\indent The fourth-order RK algorithm is obtained by defining the primitive parameters $k_1$, $k_2$, $k_3$, and $k_4$ which are calculated as: 
\begin{equation}
\begin{split}
&k_1=F(S_n) \\
&k_2=F(S_n+\frac{1}{2}h\times k_1) \\
&k_3=F(S_n+\frac{1}{2}h\times k_2) \\
&k_4=F(S_n+h\times k_3),
\end{split}
\end{equation}
where $h$ is the discrete step size. The next system state vector $S_{n+1}$ is then evaluated using the previous system state $S_n$ and these above mentioned primitive parameters. Fig. \ref{fig:HyperJerkFunc} presents the block diagram implementation of $F(S_n)$ in (\ref{eq:hyperchaosfunc}), which is reused to implement the parameters $k_i$ with different input variables. The high-level tool (XSG) automatically maps those functions to \gls{fpga} resources. The $z^{-i}$ notation indicates the insertion of $i$ registers. We can see that the longest path amounts to $5$ registers, which results in $5$ clock cycles of output latency for the function $F(S_n)$. The block diagram presented in Fig. \ref{fig:FFRK} shows the implementation of the \gls{ffrk} algorithm. The discrete step size $h=0.01$ provides enough resolution for chaotic signals. First, the parameter $k_1$ is calculated while the register enable control signal $E_1$ is high. 
\begin{equation}
S_{n+1}=\begin{bmatrix}
x_{n+1} \\ y_{n+1} \\ z_{n+1} \\ u_{n+1} \\ v_{n+1}
\end{bmatrix}
=S_n+\frac{1}{6}h\times(k_1+2k_2+2k_3+k_4). 
\end{equation}
The output $k_1$ from the block $F(S_n)$ will be temporary stored in register $R_1$. Then, the parameters $k_2$, $k_3$, and $k_4$ are produced at the high level of the control signals $E_2$, $E_3$ and $E_4$, respectively. Therefore, four primitive parameters $k_1,k_2,k_3$, and $k_4$ are stored in four registers. Finally, the next state of the system is produced when all these parameters are ready at the high period of the enable signal $E_0$ of register $R_0$. The enable signals are generated by the control unit which uses delay blocks.  
The phase spaces of the chaotic outputs of the proposed implementation shown in Fig. \ref{fig:Trajectory} are compatible with the simulation results and mathematical analysis. 
\subsection{Data Post-Processing Implementation}
\begin{figure}
	\centering
	\includegraphics[width=0.9\linewidth]{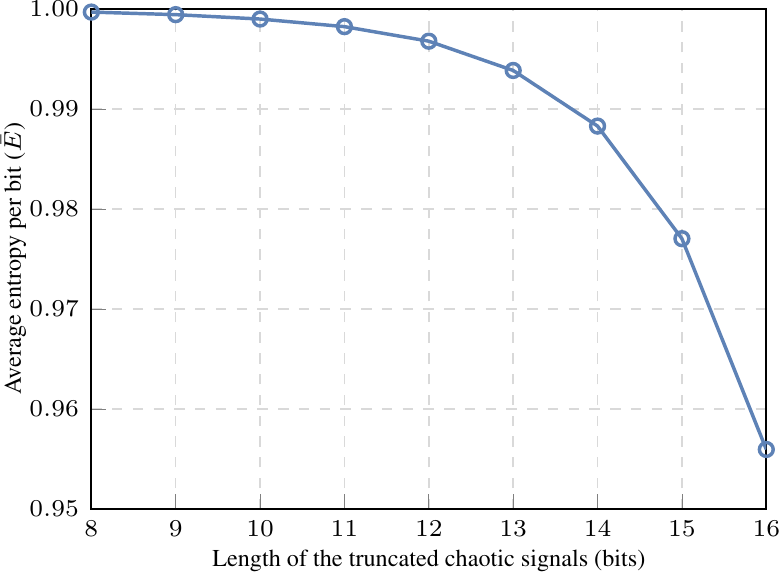}
	\caption{\centering Average entropy per bit with different length of truncated chaotic signals.}
	\label{fig:EntropyTruncated}
\end{figure}

Before we address the post-processing process, we truncate 32-bit chaotic signals as indicated in Fig. \ref{fig:PRNGScheme}. In previous works, the truncation was mentioned and applied \cite{hua_sine_2019,mao_design_2006}. In \cite{hua_sine_2019}, random bits are generated from the $33^{rd}$ bit to the $40^{th}$ bit in each 52-bit data. The authors in \cite{mao_design_2006} used 16-bits out of a 32-bit signal for the post-processing. It is obvious that the full data-width in chaotic signal is not used to generate random bits due to the low self-oscillators. Nevertheless, to obtain sufficient randomness quality, no clear reason is presented for choosing the truncation position. The least significant bits (LSBs) are preferred due to their higher fluctuation. Therefore, it is understood that the bits going from the LSB to the truncation position will be used in the post-processing. Our research shows that the signal entropy is strongly related to the truncation position. To evaluate the truncated chaotic signal's entropy, Shannon's entropy of a bitstream is calculated. For a given random variable $\Tilde{X_n}$, with a possible outcome $x_i$, $i\subset [0,N-1]$, each with probability $p_i$, the signal entropy is:
\begin{equation}
H(\Tilde{X_n})=-\sum_{i=0}^{N-1}p_i \log_{2}p_i,
\end{equation}
where $N$ is the number of symbols. The average entropy per bit is calculated as: 
\begin{equation}
\bar{E} = \frac{H(\Tilde{X_n})}{N_b}
\end{equation}
in which, $\Tilde{X_n}$ denotes the truncated chaotic signal and $N_b=\log_{2}N$ is the truncation position. 
Fig.\,\ref{fig:EntropyTruncated} provides the average entropy per bit with different lengths of truncated signals. The average entropy per bit decreases significantly when the number of remaining bits is larger than 12. Therefore, in this work, a 12-bit truncated chaotic vector $\Tilde{S_n}=[\Tilde{X_n}, \Tilde{Y_n}, \Tilde{Z_n},\Tilde{U_n},\Tilde{V_n}]$ is used in the data post-processing. 

\begin{figure}[!t]\centering
	\includegraphics[width=\linewidth]{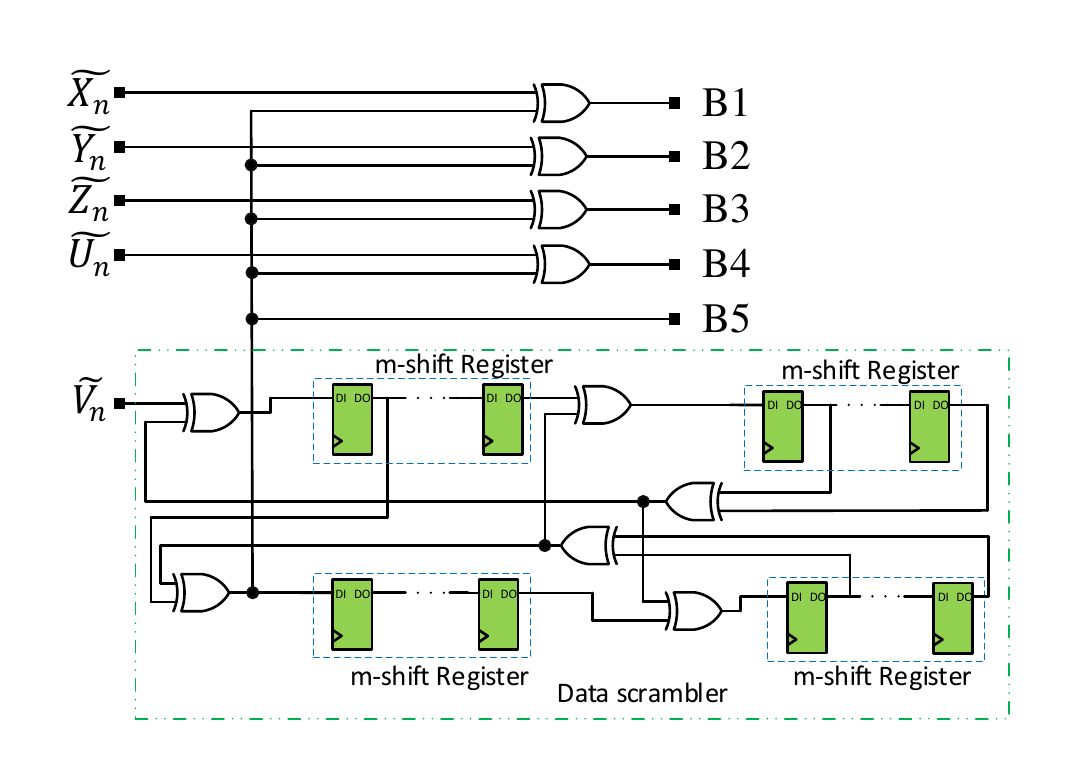}
	\caption{\centering Post-processing design. Clock signals omitted for clarity.}
	\label{fig:PP}
\end{figure}
The proposed data post-processing hardware implementation based on shift registers and XOR operations is presented in Fig. \ref{fig:PP}. The data scrambler is composed of four shift registers (m-shift registers) and XOR operations \cite{pareschi_simple_2006, rozic_iterating_2016-1}. The working principle is to evaluate the incoming bits from the truncated chaotic signals and to then reuse them. One of the chaotic signal outputs of the proposed 5D hyper-chaotic system ($\Tilde{V_n}$) is used to generate the seed for the data scrambler. Then, the other serialized truncated chaotic signals are XOR-ed with the output bit from the data scrambler. Based on the experimental result, $m=6$ shift registers, which are implemented by delay units (\glspl{ff}) in the Xilinx System Generator, are utilized to obtain high success rates in the statistical tests, .  
Five random bitstreams $(B_1\rightarrow B_5)$ are collected for the evaluation phase, which will be presented below. 
\section{Experimental Results} \label{sec:exp}
In this section, we present the deployment of the random bit generator on the Xilinx Zynq-7000 SoC ZC702 Evaluation Kit. The implementation is performed using the Xilinx System Generator tool. The random number generator is generated as an IP-core and evaluated on Xilinx Vivado tools with the default configuration. 
\subsection{Comparison and Discussion}
We start this section with a comparison with the state-of-the-art chaotic systems, i.e., a Lorenz system as well as both Elwaki's systems of \cite{elwakil_construction_2001}. 
To illustrate the advantage of our \gls{ffrk} implementation, we carried out our own implementation of the systems, in which we produced versions that either integrated the traditional RK implementation or our proposed \gls{ffrk} method. The implementation is produced on the evaluation kit mentioned above. Four iterations of the mapping functions are used to generate $k_i$ in the traditional fourth-order RK method.

\begin{figure}
	\centering
	\includegraphics[width=\linewidth]{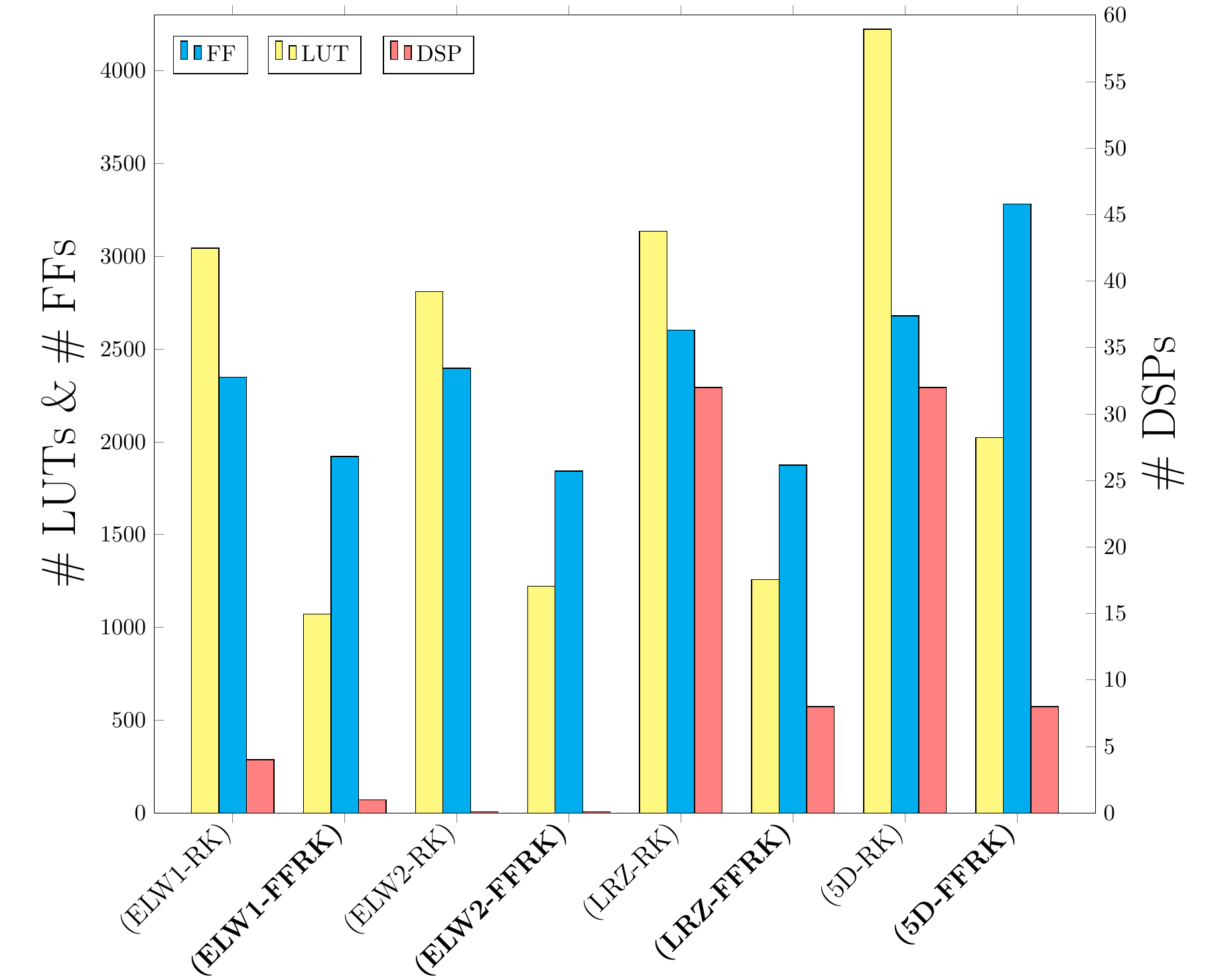}
	\caption{\centering Amount of \gls{fpga} resources in terms of \glspl{lut}, \glsentrylongpl{ff}, and DSPs required by various continuous chaotic systems, using the traditional RK implementation and the proposed \gls{ffrk}.}
	\label{fig:FFRKcomparison}
\end{figure}

Fig.\,\ref{fig:FFRKcomparison} shows the amount of \gls{fpga} resources required for the various chaotic systems. The Lorenz systems are denoted LRZ-RK and LRZ-FFRK, while the Elwaki's systems are denoted EWK1-RK, EWK2-RK, EWK1-FFRK, and EWK2-FFRK; the RK and \gls{ffrk} suffixes respectively indicate whether the system uses either the traditional RK implementation or our proposed \gls{ffrk} implementation.
From Fig.\,\ref{fig:FFRKcomparison}, it can be seen that compared to traditional RK methods \cite{azzaz_real-time_2009, zidan_effect_2011}, the proposed \gls{ffrk} requires 4 times less DSPs, under half the number of \glspl{lut}, and approximately 15\% less \glspl{ff}. 
Fig.\,\ref{fig:FFRKcomparison} also includes results for the proposed 5D chaotic system, both with the traditional RK method (denoted 5D-RK) and the proposed \gls{ffrk} method (denoted 5D-FFRK). Using the proposed \gls{ffrk} leads to a significant reduction in the amount of \glspl{lut} and DSPs required, approximately the same as that seen for the other systems. There is a small increase, ($\sim 25\%$), in the amount of \glspl{ff} required.

\begin{table*}[]
	\caption{\glsentryshort{fpga} resource comparison for various implementations of continuous chaos-based PRNGs.}
	\setlength{\tabcolsep}{4pt}
	\label{tab:Comparison}
	\centering
	\begin{tabular}{|l|c|c|c|c|c|c|c|c|c|}
		\hline
		\gls{prng} & 
		\cite{koyuncu_design_2017} &
		\multicolumn{2}{c|}{\cite{tolba_fpga_2017} }& 
		\cite{rajagopal_chaotic_2018} & \cite{rezk_reconfigurable_2019} &
		\multicolumn{2}{c|}{\cite{bonny_hardware_2019}} & 
		\cite{koyuncu_design_2020} & 
		Proposed\\[1.6ex]
		\hline
		Chaotic system & 3D & 3D (Liu) & 3D & 3D & 3D(Lorenz+Lu)  & 3D (opt. t/p) & 4D (opt. t/p) & 3D & 5D  \\
		Implementation & fourth-order RK & GL & GL & ADM & Euler & Euler & Euler & Euler & FFRK \\
		\hline 
		Slices & 12430 & 5688 & 2659 & -- & 119 & -- & -- & -- & 920 \\
		\quad \quad LUTs & 43732 & -- & -- & 1220 & 494 & 1169 & 1882 & 1355 & 2017  \\
		\quad \quad     FFs & 42092 & 4962 & 3360 & 192 & 118 & 416 & 480 & 1318 & 3458 \\
		DSPs & -- & 99 & 59 & 8 & 8 &-- & -- & -- &8\\\hline
		Max. Freq. (MHz) & 293 & 38 & 46 & 87 & 78 & 107 & 112 & 464 & 113\\
		Throughput (Mbps) & 58.7 & 1554 & 1849 & -- & 1872 &  1178 & 1869 & 464 & 6780\\\hline
	\end{tabular}
\end{table*}
In addition to the above, Table~\ref{tab:Comparison} compares our proposed hyperchaos-based PRNG implementation against the implementations of chaos-based PRNGs from the literature. In this context, we implement our system using the FFRK method using the default configurations for synthesis and implementation strategies. It is noted that changing the synthesis tool optimization policy to favor speed will lie a small improvement. However, the greatest gain will come from duplicating the design along with some straightforward circuitry to merge their results, effectively doubling the throughput at the cost of doubling the amount of required resources without any effect on latency.

In terms of resource requirements, it can be observed that the 3D chaotic system using the fourth-order RK algorithm in \cite{koyuncu_design_2017}, which utilized up to 32\% of the available resources from the Virtex-6 XC6VLX240T-1-FF1156 FPGA chip, requires more resources than the other systems. Compared to our work, the Gr\"{u}nwald-Letnikov (GL)-based algorithms of \cite{tolba_fpga_2017} require a similar number of \glspl{ff} and from $3\times$ to $6.4\times$ the number of slices. While the implementation of the 3D-fractional-order chaotic system using the Adomian decomposition method (ADM) from \cite{rajagopal_chaotic_2018} requires a smaller number of slices, the random number generator is only performed in software design. The 3D-chaos based PRNG, which is itself based on Lorenz and Lu chaotic systems from \cite{rezk_reconfigurable_2019} using the Euler method, uses the least number of resources among all systems, but offers an achievable throughput of $1872$ Mbps, which is much lower than that of our proposed system. The 4D-chaos based system from \cite{bonny_hardware_2019}, which is optimized for throughput, is the one that comes the closest to our system in terms of the number of LUTs, where our implementation achieves $3.6\times$ greater throughput. The system in \cite{koyuncu_design_2020} uses the highest number of resources among all the 3D-chaos based PRNG implementations using the Euler method. We recall that the hardware implementation of the Euler method requires fewer device resources, but achieves less accuracy as compared to the fourth-order Runge-Kutta algorithm \cite{zidan_effect_2011}.

Our design seeks to reach a high throughput, and a high level of randomness while requiring a modest amount of resources. The Euler method implementation provides less accuracy, and therefore the average entropy per bit is decreased and so is the randomness level for the output random bits. 
We also note that for the targeted FPGA, the proposed system requires under 4\% of the available LUTs, FFs, or DSP blocks. The design works at a maximum frequency of 113 MHz, which enables a maximum throughput of 6.78 Gbps for the generator. The estimated power consumption of the random number generator is 73mW@6.78Gbps. Therefore, the estimated energy efficiency is 10.8 pJ/b. The \gls{ffrk} implementation for the chaotic system has a delay of 65 clock cycles while there is only one latency at the random outputs. 

\begin{figure*}
	\centering
	\includegraphics[width=\linewidth]{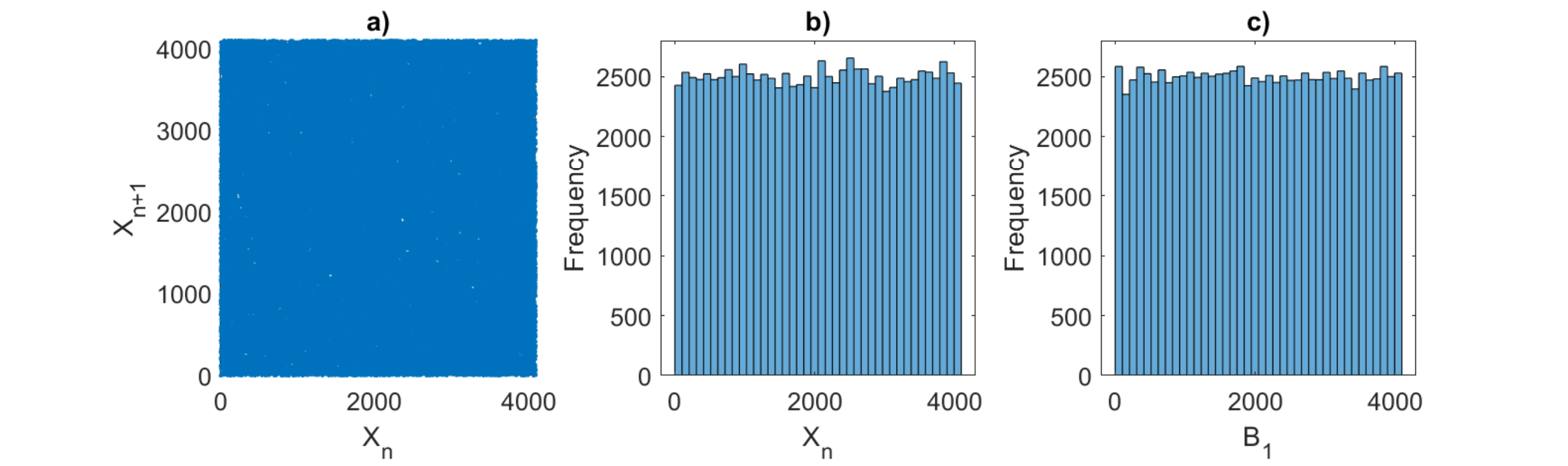}
	\caption{12b-truncated chaotic signals in a) X-Y distribution points b) Histogram diagram. Before post-processing and c) Histogram of output bitstreams - After post-processing.}
	\label{fig:TruncatedHist}
\end{figure*}
\section{Randomness Performance Evaluation}\label{sec:perf-eval}
\begin{figure}
	\centering
	\includegraphics[width=0.8\linewidth]{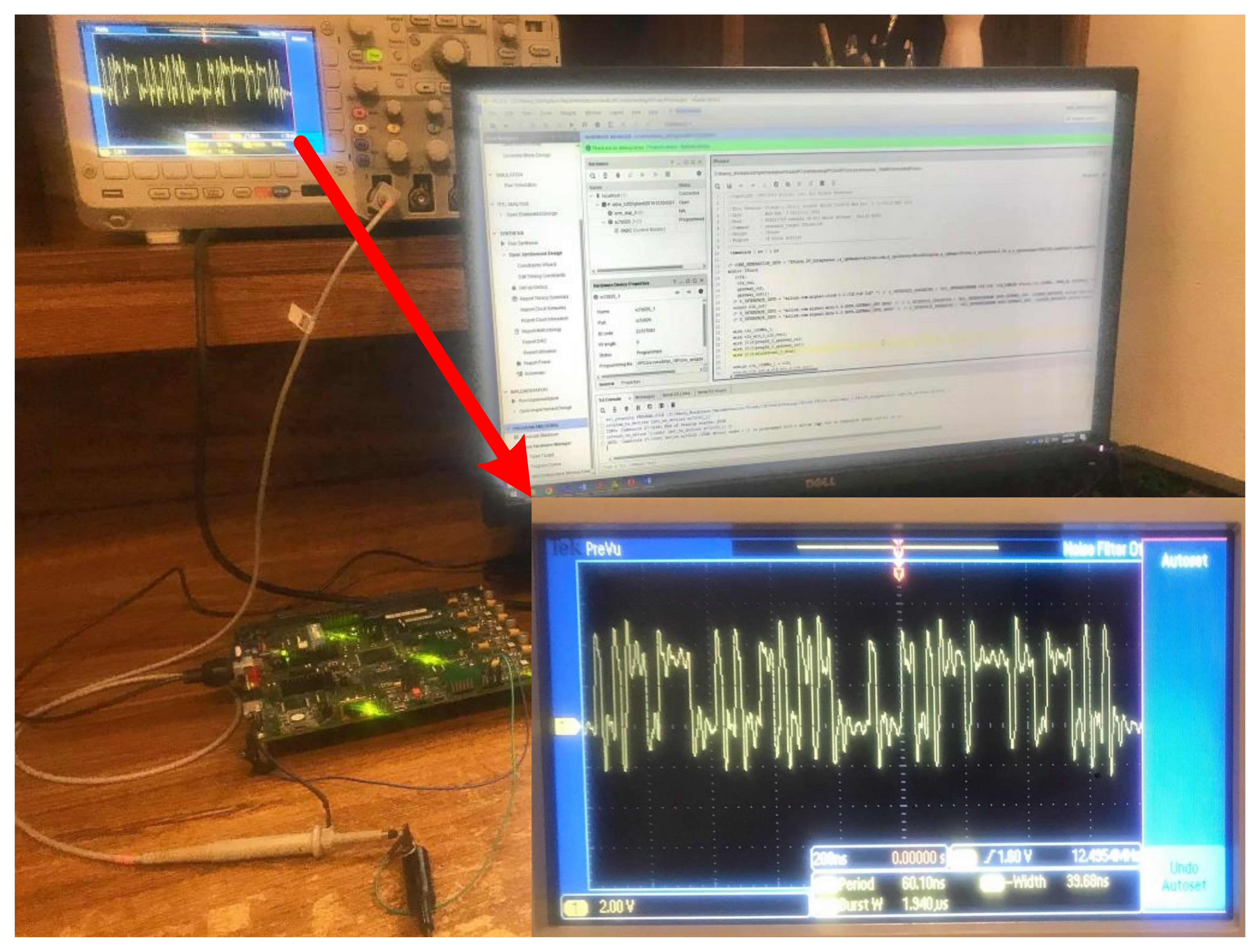}
	\caption{\centering The experimental validation of the proposed chaos-based PRNG at the 50MHz clock frequency.}
	\label{fig: Setup}
\end{figure}

In this section, we present multiple tests performed on the harvested binary bitstreams before and after applying post processing. These bitstreams are collected by hardware co-simulation of Vivado System Generator and Matlab. The PRNG core runs on the FPGA board connected to the computer and the output data are collected through JTAG. The experimental setup is illustrated in Fig.\,\ref{fig: Setup}. The random binary outputs are observed using an oscilloscope.
\subsection{Histogram}
First, $100K$ length of 12b-truncated chaotic signals are collected and used in distribution and histogram analysis. As shown in Fig. \ref{fig:TruncatedHist}-a), since there are no obvious patterns in their distribution, 12b-truncated chaotic signals have good randomness characteristics. Fig. \ref{fig:TruncatedHist}-b) shows the histogram of the signal $\Tilde{X_n}$ before post-processing where several peaks that can be observed. The histogram of the output bitstream $B_1$ after post-processing is illustrated in Fig. \ref{fig:TruncatedHist}-c), which indicates a random distribution, where no apparent pattern can be discerned. Similar results are observed for $B_2$ to $B_5$.
\subsection{Standard statistical random testsuites}
In order to be used in cryptography applications, the random binary bits should be tested using statistical tests which require long bitstreams. 
The NIST SP800-22 test-suite, which is developed and introduced by The National Institute of Standards and Technology (NIST), is commonly used \cite{NIST-SP800-22:2010}. The NIST test results are presented in Table II. Moreover, Table III provides the results for the Diehard tests which consist of 12 sub-tests. TestU01 is another commonly used test-suite providing a variety of statistical tests for random bit generators. We applied two battery tests including the Rabbit and the Alphabit tests, to the binary sequences from the proposed random bit generator. The Rabbit test includes 39 subtests and the Alphabit test includes 17 subtests for a bitstream of $2^{25}$ bit length.
The battery test results are presented in Table \ref{tab:TestU01}. Most of the tests are passed, except for the multinomial test of bitstream $B_3$, with a P-value = 0.00097, which is very close to the threshold (0.001). It should not be understood that if a RNG fails in some statistical tests, it cannot be used in practical problems; even the best commercial RNGs experience failures in very complicated tests \cite{TestU01-2007}.  
\begin{table}
	\caption{\centering NIST SP800-22 tests for the proposed PRNG, 
		$N=10^3$ sequences of the length of 1Mb.}
	\label{tab:NISTTest}
	\centering
	\begin{tabular}{|l|l| l|}
		\hline \\[-3mm]
		NIST SP-800.22 & P-value(*) & Proportion  \\
		\hline
		
		Monobit test & 	0.323668  & 998/1000 \\
		Frequency within block test & 0.763677  & 997/1000 \\
		Runs test & 0.164425 & 992/1000  \\
		Longest run 1's test  & 0.630872       & 1000/1000 \\
		Rank test & 0.575608      & 990/1000  \\
		DFT test & 0.655608        & 996/1000 \\
		Cumulative sum & 0.924844       & 997/1000  \\
		Overlapping template & 0.130366      & 990/1000 \\
		Non-overlapping template & 0.999999  & 997/1000  \\  
		Linear complexity test & 0.663130      & 998/1000 \\
		Maurers universal test & 0.999142      & 998/1000 \\
		Approximate entropy & 0.956970      & 1000/1000 \\
		Serial & 0.740523      & 1000/1000 \\
		Random excursions & 0.304039      & 997/1000 \\
		Random excursion variant & 0.049499     & 988/1000 \\ 
		\hline
	\end{tabular}
	\\[2mm]
	(*): the average value. 
\end{table}
\begin{table}
	\caption{\centering Diehard tests for the proposed PRNG, sequence of 5M bits.}
	\label{tab:DiehardTest}
	\centering
	\begin{tabular}{|l| l| l|}
		\hline \\[-3mm]
		Dieharder & P-value & Result\\
		\hline
		Birthdays & 	0.0528  & PASSED \\
		OPERM5 & 0.0000  & FAILED \\
		Rank 6x8  & 0.0042       & WEAK \\
		Bitstream & 0.9566 & PASSED \\
		OPSO & 0.0109 & PASSED \\
		DNA & 0.1829 & PASSED \\ 
		Count 1s string & 0.9989 & PASSED \\
		Parking lot & 0.2452 & PASSED \\ 
		2d sphere & 0.1491 & PASSED \\
		3d sphere & 0.1581 & PASSED \\
		Sums & 0.0222 & PASSED \\
		Runs & 0.9288 & PASSED \\
		Craps & 0.4094 & PASSED \\ 
		\hline
	\end{tabular}
\end{table}
\begin{table}
	\caption{Battery test results for $2^{25}$ output binary bits.}
	\label{tab:TestU01}
	\centering
	\begin{tabular}{|l| c| c|}
		\hline \\[-3mm]
		TestU01 & Rabbit & Alphabit \\\hline
		$B_1$ & 39/39 & 17/17 \\
		$B_2$ & 39/39 & 17/17 \\
		$B_3$ & 38/39 & 17/17 \\
		$B_4$ & 39/39 & 17/17 \\
		$B_5$ & 39/39 & 17/17 \\
		\hline 
	\end{tabular}
\end{table}
\section{Conclusion}\label{sec:conclusion}
We have presented a novel 5D hyperchaotic system along with its implementation on \gls{fpga}. In this work, we proposed a novel method, which reduces device resource usage, to implement the fourth-order Runge-Kutta numerical technique to solve differential equations in chaotic systems. The PRNG, which is designed to take advantage of the high dimension and the high accuracy of the hyperchaotic system's hardware implementation, reaches a maximum throughput of 6.78\, Gbps. The design, which is deployed in the Xilinx FPGA chip, occupies under 4\% of the available \gls{fpga} hardware resources. The random generated bits have been verified in modern statistical tests confirming that our proposed system is suitable for industrial applications. Future work includes integrating the proposed PRNG in real-time communication applications such as the secure streaming of images and videos.



\bibliographystyle{IEEEtranTIE}
\bibliography{ALL_xx-TIE-xxxx}\ 

\begin{thebibliography}{10}
\providecommand{\url}[1]{#1}
\csname url@samestyle\endcsname
\providecommand{\newblock}{\relax}
\providecommand{\bibinfo}[2]{#2}
\providecommand{\BIBentrySTDinterwordspacing}{\spaceskip=0pt\relax}
\providecommand{\BIBentryALTinterwordstretchfactor}{4}
\providecommand{\BIBentryALTinterwordspacing}{\spaceskip=\fontdimen2\font plus
\BIBentryALTinterwordstretchfactor\fontdimen3\font minus
  \fontdimen4\font\relax}
\providecommand{\BIBforeignlanguage}[2]{{%
\expandafter\ifx\csname l@#1\endcsname\relax
\typeout{** WARNING: IEEEtran.bst: No hyphenation pattern has been}%
\typeout{** loaded for the language `#1'. Using the pattern for}%
\typeout{** the default language instead.}%
\else
\language=\csname l@#1\endcsname
\fi
#2}}
\providecommand{\BIBdecl}{\relax}
\BIBdecl

\bibitem{li_novel_2019}
B.~Li, X.~Liao, and Y.~Jiang, ``\BIBforeignlanguage{en}{A novel image
  encryption scheme based on improved random number generator and its
  implementation},'' \emph{\BIBforeignlanguage{en}{Nonlinear Dyn.}}, vol.~95,
  no.~3, pp. 1781--1805, Feb. 2019.

\bibitem{liu_cryptanalysis_2020}
Y.~Liu, Z.~Qin, X.~Liao, and J.~Wu, ``\BIBforeignlanguage{en}{Cryptanalysis and
  enhancement of an image encryption scheme based on a 1-{D} coupled {Sine}
  map},'' \emph{\BIBforeignlanguage{en}{Nonlinear Dyn.}}, vol. 100, no.~3, pp.
  2917--2931, May. 2020.

\bibitem{moysis_chaos_2020}
L.~Moysis, A.~Tutueva, C.~Volos, and D.~Butusov, ``\BIBforeignlanguage{en}{A
  {Chaos} {Based} {Pseudo}-{Random} {Bit} {Generator} {Using} {Multiple}
  {Digits} {Comparison}},'' \emph{\BIBforeignlanguage{en}{Chaos Theory and
  Applications}}, vol.~2, no.~2, p.~11, Jul. 2020.

\bibitem{NN-Access}
N.~{Nguyen}, L.~{Pham-Nguyen}, M.~B. {Nguyen}, and G.~{Kaddoum}, ``A low power
  circuit design for chaos-key based data encryption,'' \emph{IEEE Access},
  vol.~8, pp. 104\,432--104\,444, 2020.

\bibitem{liu_hyperchaotic_2016}
Y.~Liu and X.~Tong, ``\BIBforeignlanguage{en}{Hyperchaotic system-based
  pseudorandom number generator},'' \emph{\BIBforeignlanguage{en}{IET
  Information Security}}, vol.~10, no.~6, pp. 433--441, Nov. 2016.

\bibitem{irfan_pseudorandom_2020}
M.~Irfan, A.~Ali, M.~A. Khan, M.~Ehatisham-ul Haq, S.~N. Mehmood~Shah,
  A.~Saboor, and W.~Ahmad, ``\BIBforeignlanguage{en}{Pseudorandom {Number}
  {Generator} ({PRNG}) {Design} {Using} {Hyper}-{Chaotic} {Modified} {Robust}
  {Logistic} {Map} ({HC}-{MRLM})},''
  \emph{\BIBforeignlanguage{en}{Electronics}}, vol.~9, no.~1, p. 104, Jan.
  2020.

\bibitem{zidan_effect_2011}
M.~A. Zidan, A.~G. Radwan, and K.~N. Salama, ``\BIBforeignlanguage{en}{The
  effect of numerical techniques on differential equation based chaotic
  generators},'' in \emph{\BIBforeignlanguage{en}{{ICM} 2011 {Proceeding}}},
  pp. 1--4, Hammamet, Tunisia, Dec. 2011.

\bibitem{akgul_chaos-based_2016}
A.~Akgul, H.~Calgan, I.~Koyuncu, I.~Pehlivan, and A.~Istanbullu,
  ``\BIBforeignlanguage{en}{Chaos-based engineering applications with a {3D}
  chaotic system without equilibrium points},''
  \emph{\BIBforeignlanguage{en}{Nonlinear Dyn.}}, vol.~84, no.~2, pp. 481--495,
  Apr. 2016.

\bibitem{koyuncu_design_2017}
I.~Koyuncu and A.~Turan~{\"O}zcerit, ``\BIBforeignlanguage{en}{The design and
  realization of a new high speed {FPGA}-based chaotic true random number
  generator},'' \emph{\BIBforeignlanguage{en}{Computers \& Electrical
  Engineering}}, vol.~58, pp. 203--214, Feb. 2017.

\bibitem{zhang_system_2017}
L.~Zhang, ``\BIBforeignlanguage{en}{System generator model-based {FPGA} design
  optimization and hardware co-simulation for {Lorenz} chaotic generator},'' in
  \emph{\BIBforeignlanguage{en}{2017 2nd {Asia}-{Pacific} {Conf.} on
  {Intelligent} {Robot} {Systems} ({ACIRS})}}, pp. 170--174.\hskip 1em plus
  0.5em minus 0.4em\relax Wuhan, China: IEEE, Jun. 2017.

\bibitem{tolba_fpga_2017}
M.~F. Tolba, A.~M. AbdelAty, N.~S. Soliman, L.~A. Said, A.~H. Madian, A.~T.
  Azar, and A.~G. Radwan, ``\BIBforeignlanguage{en}{{FPGA} implementation of
  two fractional order chaotic systems},'' \emph{\BIBforeignlanguage{en}{AEU -
  Int J Electron Commun}}, vol.~78, pp. 162--172, Aug. 2017.

\bibitem{koyuncu_analog_2013}
I.~Koyuncu, A.~T. Ozcerit, and I.~Pehlivan, ``\BIBforeignlanguage{en}{An analog
  circuit design and {FPGA}-based implementation of the {Burke}-{Shaw} chaotic
  system},'' \emph{\BIBforeignlanguage{en}{Optoelectron. Adv. Mater. Rapid
  Commun.}}, p.~5, 2013.

\bibitem{tlelo-cuautle_fpga_2015}
E.~Tlelo-Cuautle, J.~Rangel-Magdaleno, A.~Pano-Azucena, P.~Obeso-Rodelo, and
  J.~Nunez-Perez, ``\BIBforeignlanguage{en}{Fpga realization of multi-scroll
  chaotic oscillators},'' \emph{\BIBforeignlanguage{en}{Communications in
  Nonlinear Science and Numerical Simulation}}, vol.~27, no. 1-3, pp. 66--80,
  Oct. 2015.

\bibitem{avaroglu_hybrid_2015}
E.~Avaroglu, I.~Koyuncu, A.~B. Ozer, and M.~Turk,
  ``\BIBforeignlanguage{en}{Hybrid pseudo-random number generator for
  cryptographic systems},'' \emph{\BIBforeignlanguage{en}{Nonlinear Dyn.}},
  vol.~82, no. 1-2, pp. 239--248, Oct. 2015.

\bibitem{bonny_hardware_2019}
T.~Bonny, R.~Al~Debsi, S.~Majzoub, and A.~S. Elwakil,
  ``\BIBforeignlanguage{en}{Hardware {Optimized} {FPGA} {Implementations} of
  {High}-{Speed} {True} {Random} {Bit} {Generators} {Based} on
  {Switching}-{Type} {Chaotic} {Oscillators}},''
  \emph{\BIBforeignlanguage{en}{Circuits, Systems, and Signal Processing}},
  vol.~38, no.~3, pp. 1342--1359, Mar. 2019.

\bibitem{akgul_design_2019}
A.~Akgul, C.~Arslan, and B.~Aricioglu, ``\BIBforeignlanguage{en}{Design of an
  {Interface} for {Random} {Number} {Generators} based on {Integer} and
  {Fractional} {Order} {Chaotic} {Systems}},''
  \emph{\BIBforeignlanguage{en}{Chaos Theory and Applications}}, vol.~1, no.~1,
  pp. 1--18, Nov. 2019, number: 1.

\bibitem{barati_simple_2016}
K.~Barati, S.~Jafari, J.~C. Sprott, and V.-T. Pham, ``Simple chaotic flows with
  a curve of equilibria,'' \emph{Int J Bifurcat Chaos}, vol.~26, no.~12, p.
  1630034, Nov. 2016.

\bibitem{fozin_fonzin_coexisting_2018}
T.~Fozin~Fonzin, K.~Srinivasan, J.~Kengne, and F.~Pelap,
  ``\BIBforeignlanguage{en}{Coexisting bifurcations in a memristive
  hyperchaotic oscillator},'' \emph{\BIBforeignlanguage{en}{AEU - Int J
  Electron Commun}}, vol.~90, pp. 110--122, Jun. 2018.

\bibitem{bao_initial_2018}
H.~Bao, N.~Wang, B.~Bao, M.~Chen, P.~Jin, and G.~Wang, ``Initial
  condition-dependent dynamics and transient period in memristor-based
  hypogenetic jerk system with four line equilibria,'' \emph{Communications in
  Nonlinear Science and Numerical Simulation}, vol.~57, pp. 264--275, Apr.
  2018.

\bibitem{cavusoglu_new_2019}
{\"U}.~{\c{C}}avu{\c{s}}o{\u{g}}lu, S.~Panahi, A.~Akg{\"u}l, S.~Jafari, and
  S.~Kacar, ``\BIBforeignlanguage{en}{A new chaotic system with hidden
  attractor and its engineering applications: analog circuit realization and
  image encryption},'' \emph{\BIBforeignlanguage{en}{Analog Integr Circ Sig
  Process}}, vol.~98, no.~1, pp. 85--99, Jan. 2019.

\bibitem{pham_chaotic_2016}
V.-T. Pham, S.~Vaidyanathan, C.~K. Volos, S.~Jafari, and X.~Wang, ``A chaotic
  hyperjerk system based on memristive device,'' vol. 636, p.~21.

\bibitem{chlouverakis_comparison_2005}
K.~E. Chlouverakis and J.~Sprott, ``\BIBforeignlanguage{en}{A comparison of
  correlation and {Lyapunov} dimensions},''
  \emph{\BIBforeignlanguage{en}{Physica D: Nonlinear Phenomena}}, vol. 200, no.
  1-2, pp. 156--164, Jan. 2005.

\bibitem{hua_sine_2019}
Z.~Hua, B.~Zhou, and Y.~Zhou, ``\BIBforeignlanguage{en}{Sine {Chaotification}
  {Model} for {Enhancing} {Chaos} and {Its} {Hardware} {Implementation}},''
  vol.~66, no.~2, pp. 1273--1284, Feb. 2019.

\bibitem{mao_design_2006}
Y.~Mao, L.~Cao, and W.~Liu, ``\BIBforeignlanguage{en}{Design and {FPGA}
  {Implementation} of a {Pseudo}-{Random} {Bit} {Sequence} {Generator} {Using}
  {Spatiotemporal} {Chaos}},'' in \emph{\BIBforeignlanguage{en}{2006 {Int.}
  {Conf.} on {Communications}, {Circuits} and {Systems}}}, pp.
  2114--2118.\hskip 1em plus 0.5em minus 0.4em\relax Guilin, Guangzi, China:
  IEEE, Jun. 2006.

\bibitem{pareschi_simple_2006}
F.~Pareschi, R.~Rovatti, and G.~Setti, ``Simple and effective post-processing
  stage for random stream generated by a chaos-based {RNG},'' in \emph{The 2006
  Int. Symp. on Nonlinear Theory and its Applications ({NOLTA}2006)}, p.~5,
  2006.

\bibitem{rozic_iterating_2016-1}
V.~Rozic, B.~Yang, W.~Dehaene, and I.~Verbauwhede, ``Iterating von neumann's
  post-processing under hardware constraints,'' in \emph{2016 {IEEE} Int. Symp.
  on Hardware Oriented Security and Trust ({HOST})}, pp. 37--42.\hskip 1em plus
  0.5em minus 0.4em\relax {IEEE}, 2016-05.

\bibitem{elwakil_construction_2001}
A.~Elwakil and M.~Kennedy, ``\BIBforeignlanguage{en}{Construction of classes of
  circuit-independent chaotic oscillators using passive-only nonlinear
  devices},'' vol.~48, no.~3, pp. 289--307, Mar. 2001.

\bibitem{azzaz_real-time_2009}
M.~S. Azzaz, C.~Tanougast, S.~Sadoudi, and A.~Dandache,
  ``\BIBforeignlanguage{en}{Real-time {FPGA} implementation of {Lorenz}'s
  chaotic generator for ciphering telecommunications},'' in
  \emph{\BIBforeignlanguage{en}{2009 {Joint} {IEEE} {North}-{East} {Workshop}
  on {Circuits} and {Systems} and {TAISA} {Conf.}}}, pp. 1--4.\hskip 1em plus
  0.5em minus 0.4em\relax Toulouse, France: IEEE, Jun. 2009.

\bibitem{rajagopal_chaotic_2018}
K.~Rajagopal, A.~Akgul, S.~Jafari, and B.~Aricioglu,
  ``\BIBforeignlanguage{en}{A chaotic memcapacitor oscillator with two unstable
  equilibriums and its fractional form with engineering applications},''
  \emph{\BIBforeignlanguage{en}{Nonlinear Dyn}}, vol.~91, no.~2, pp. 957--974,
  Jan. 2018.

\bibitem{rezk_reconfigurable_2019}
A.~A. Rezk, A.~H. Madian, A.~G. Radwan, and A.~M. Soliman,
  ``\BIBforeignlanguage{en}{Reconfigurable chaotic pseudo random number
  generator based on {FPGA}},'' \emph{\BIBforeignlanguage{en}{AEU - Int J
  Electron Commun.}}, vol.~98, pp. 174--180, Jan. 2019.

\bibitem{koyuncu_design_2020}
I.~Koyuncu, M.~Tuna, I.~Pehlivan, C.~B. Fidan, and M.~Alcin,
  ``\BIBforeignlanguage{en}{Design, {FPGA} implementation and statistical
  analysis of chaos-ring based dual entropy core true random number
  generator},'' \emph{\BIBforeignlanguage{en}{Analog Integr Circ Sig Process}},
  vol. 102, no.~2, pp. 445--456, Feb. 2020.

\bibitem{NIST-SP800-22:2010}
A.~Rukhin \emph{et~al.}, \emph{A statistical test suite for random and
  pseudorandom number generators for cryptographic applications}, {National
  Institute of Standards and Technology (NIST)} Special Publication 800-22,
  Rev.~{1a}, Apr. 2010.

\bibitem{TestU01-2007}
P.~L'Ecuyer and R.~Simard, ``Testu01: A c library for empirical testing of
  random number generators,'' \emph{ACM Trans. Math. Softw.}, vol.~33, no.~4,
  Aug. 2007.

\end{thebibliography}

\end{document}